\documentclass[aps,11 pt]{revtex4}
\usepackage{amssymb}
\usepackage{amsmath}
\usepackage{graphicx}
\usepackage{epsfig,amsmath}
\usepackage[bookmarksnumbered,linktocpage,pdfstartview=FitH]{hyperref}
\usepackage{longtable}
\usepackage{bm}
\usepackage{morefloats}

\begin{document}

\title{Fluctuations of the vortex line density in turbulent flows of quantum
fluids}
\author{Sergey K. Nemirovskii\thanks{%
email address: nemir@itp.nsc.ru}}
\affiliation{Institute of Thermophysics, Lavrentyev ave, 1, 630090, Novosibirsk, Russia,\\
Novosibirsk State University, Novosibirsk Russia}
\date{\today }

\begin{abstract}
We present an analytical study of fluctuations of the Vortex Line Density
(VLD) $<\delta \mathcal{L}(\omega )\;\delta \mathcal{L}(-\omega )>$ in
turbulent flows of quantum fluids. Two cases are considered. The first one
is the counterflowing (Vinen) turbulence, where the vortex lines are
disordered, and the evolution of quantity $\mathcal{L}(t)$ obeys the Vinen
equation. The second case is the quasi-classic turbulence, where vortex
lines are believed to form the so called vortex bundles, and their dynamics
is described by the HVBK equations. The latter case, is of a special
interest, since a number of recent experiments demonstrate the $\omega
^{-5/3}$ dependence for spectrum VLD, instead of $\omega ^{1/3}$ law,
typical for spectrum of vorticity. In steady states the VLD \ is related to
the normal velocity \ as $\mathcal{L=}~(\rho \gamma /\rho
_{s})^{2}\;v_{n}^{2}$ for the Vinen case, and $\mathcal{L}=\left\vert \nabla
\times \mathbf{v}_{n}\right\vert /\kappa $ for rotating vortex tubes. In
nonstationary situation, in particular, in the fluctuating turbulent flow
there is a retardation between the instantaneous value of the normal velocity
and the quantity $\mathcal{L}$. This retardation tends to decrease in
the accordance with the inner dynamics, which has a relaxation character. In both
cases the relaxation dynamics of VLD is related to fluctuations of the
relative velocity, however if for the Vinen case the rate of temporal change
for $\mathcal{L}(t)$ is directly depends on $\delta \mathbf{v}_{ns}$, for
the HVBK dynamics it depends on $\nabla \times \delta \mathbf{v}_{ns}$. As a
result, for the disordered case the spectrum $<\delta \mathcal{L}(\omega
)\;\delta \mathcal{L}(-\omega )>$ coincides with the spectrum $\omega
^{-5/3} $. In the case of the bundle arrangement, the spectrum of the VLD
varies (at different temperatures) from $\omega ^{1/3}$ to $\omega ^{-5/3}$
dependencies. This conclusion may serve as a basis for  the experimental
determination of what kind of the turbulence is implemented in different
types of generation.

PACS number(s): 67.25.dk, 47.37.+q, 03.75.Kk
\end{abstract}

\maketitle

\section{Introduction}

The problem of modeling classical turbulence with a set of chaotic quantized
vortices is the hottest topic in modern studies of vortex tangles in quantum
fluids (see e.g., recent reviews articles \cite{Vinen2010},\cite%
{Procaccia2008},\cite{Skrbek2012}). The most common view of quasi-classical
turbulence is the model of vortex bundles. The point is that the quantized
vortices have a fixed core radius, so they do not possess the important
property of classical turbulence -- stretching of tubes -- which is
responsible for the turbulence energy cascade from the large scales to the
small scales. The collections of the near-parallel quantized vortices
(vortex bundles) do possess this property, so the idea that quasi-classical
turbulence in quantum fluids is realized via vortex bundles of different
sizes and intensities (number of threads ) seems to be quite natural.

Recently two numerical evidences of the vortex bundles structures were
obtained. Thus, in one numerical work, \cite{Sasa2011}, in $2048^{3}$
simulation of quantum turbulence within the Gross-Pitaevskii equation,
authors observed nonuniform structures. The authors claimed that "the
visualization of vortices clearly shows the bundle-like structure, which has
never been confirmed in GPE simulations on smaller grids." In the other
numerical work \cite{Baggaley2011a},\cite{Baggaley2012a}, the authors
studied the evolution of the vortex structures (at zero temperature) on the
basis of the Biot-Savart law. They also observed structures reminiscent of
the field of vorticity in classical turbulence (see e.g., \cite{Vincent1991}%
).

As for experimental confirmation, there are not so far strong evidences of
the bundle structure. On the contrary, there are experimental results, which
would seem to refute the idea of bundles. Thus, in experiments by Roche et
al. \cite{Roche2007}, \ and by Bradley et al.\cite{Bradley2008},\ it was
observed that the spectrum of the fluctuation of the VLD $\mathcal{L}$ is
compatible with a $-5/3$ power law. This contradicts the idea of the bundles
structure, since the spectrum of the vorticity (and, correspondingly, of the
VLD $\mathcal{L}$ (via Eq. (\ref{bundle omega}))) should scale as $1/3$
power law. An explanation was offered by Roche and Barenghi \cite{Roche2008}%
. The authors considered the VLD $\mathcal{L}$ to be decomposed into two
components. The one,, smaller, part consisting of the polarized component,
is responsible for the large scale turbulent phenomena, whereas the other,
disordered, part evolves as a "passive" scalar, thereby taking the $-5/3$
velocity spectrum. In paper \cite{Salort2011}, the authors performed the
direct numerical simulations of the "truncated HVBK" model. They confirmed
the existence of the $\omega ^{-5/3}$ spectrum for fluctuation of the VLD,
however, for the larger temperature this dependence became more shallow
(probably reaching $\omega ^{1/3}$), as it should be for the classic
turbulence.

We present an analytical evaluation of the spectrum of fluctuations VLD $%
<\delta \mathcal{L}(\omega )\;\delta \mathcal{L}(-\omega )>$. Two cases are
considered. The first one is the counterflowing (Vinen) turbulence, where
vortex lines are disordered and dynamics of quantity $\mathcal{L}(t)$ is
governed by the Vinen equation. The second case is the quasi-classic
turbulence, where the vortex lines are believed to form the so called vortex
bundles, and the dynamics of VLD obeys the HVBK equations. In steady states
the VLD \ is related to the normal velocity \ as $\mathcal{L=}~(\rho \gamma
/\rho _{s})^{2}\;v_{n}^{2}$ for the Vinen case and $\mathcal{L}=\left\vert
\nabla \times \mathbf{v}_{n}\right\vert /\kappa $ for the rotating vortex
tubes (notations are standard, see e.g. \cite{Donnelly1991}). In
nonstationary situation, in particular, in the fluctuating turbulent flow
there is a retardation between instantaneous value of the normal velocity
and the quantity $\mathcal{L}(t)$. This retardation tends to decrease,
according to the inner dynamics, which has a relaxation character. In both
cases, the relaxation of $\delta \mathcal{L}(t)$ is related to fluctuations
of the normal velocity $\delta \mathbf{v}_{ns}$ velocity. If, however, for
the Vinen disordered turbulence, the rate of temporal change for $\mathcal{L}%
(t)$ is directly depends on $\delta \mathbf{v}_{ns}$, for the HVBK dynamics
it depends on the quantity $\nabla \times \delta \mathbf{v}_{ns}$. In
addition, the relaxation mechanisms, and, consequently, times of relaxation
are different. The factors, outlined above lead to different formulas for
spectra $<\delta \mathcal{L}(\omega )\;\delta \mathcal{L}(-\omega )>$ and
their dependence on temperature.

\section{Vinen Equation case}

Let us study reaction of the vortex line density, in fluctuating flow of
normal velocity supposing that the dynamics of $\mathcal{L}(t)$ obeys the
Vinen equation

\begin{equation}
\frac{\partial \mathcal{L}}{\partial t}=~\alpha _{V}\;|\mathbf{v}_{ns}|\;%
\mathcal{L}^{3/2}\;-\;\beta _{V}\;\mathcal{L}^{2}.  \label{VE 1}
\end{equation}%
Equation (\ref{VE 1}) was derived phenomenologically initially for pure
counterflowing superfluid helium \cite%
{Vinen1957,Vinen1957b,Vinen1957c,Vinen1958}. Attempts to derive it an
analytic form \cite{Schwarz1978,Schwarz1988,Nemirovskii2007a,Nemirovskii2008}
demonstrated that this equation is seemingly valid for any non-structured
turbulence. Under term "non-structured turbulence" we understand the vortex
tangle, which consists of closed vortex loops of different sizes, uniformly
distributed in space. It differs, for instanse, from the turbulent fronts in
rotating fluids, which deals with the lines terminating on lateral walls. It
also differs from the the mechanically excited turbulence, which is believed
to consists of the so called vortex bundles composed of very polarized
vortex filaments.  Beside of the counterflow turbulence, the "non-structured
turbulence" is generated by intensive sounds (both by the first and second).
The case of vortex tangles. which appear also after the quench due to the
Kibble-Zurek mechanism, or by the proliferation of vortices when approaching
the critical temperature.

Our goal now is to study the stochastic properties of $\mathcal{L}(t)$, when
$\mathbf{v}_{n}$ fluctuates with the a given spectral density $\left\langle
\delta \mathbf{v}_{n}(\omega )\delta \mathbf{v}_{n}(-\omega )\right\rangle
=f(\omega )$. Further, for simplicity, we will study the pure counterflowing
case in sense that the net flow is absent, $\mathbf{j=}\rho _{n}\mathbf{v}%
_{n}+\rho _{s}\mathbf{v}_{s}=0$. Then, the average value $\mathcal{L}_{0}$of
the vortex line density\ is related to a relative velocity $\mathbf{v}_{ns}=%
\mathbf{v}_{n}-\mathbf{v}_{n}$ by the usual relation
\begin{equation}
\mathcal{L}_{0}~=~{\frac{\alpha _{V}^{2}}{{\beta _{V}^{2}}}}\;\mathbf{v}%
_{ns}^{2}~=~\gamma ^{2}\;\mathbf{v}_{ns}^{2}.  \label{VLD v squared}
\end{equation}%
To take into account the fluctuations, we put
\begin{equation}
\mathcal{L=}~\gamma ^{2}\;v_{ns}^{2}+\delta \mathcal{L},\ \ \ \ \ \ \
\mathbf{v}_{n}=\mathbf{v}_{n0}+\delta \mathbf{v}_{n}.  \label{linearization}
\end{equation}%
From condition $\mathbf{j=}0$ it follows
\begin{equation}
\mathbf{v}_{s}=\mathbf{v}_{s0}-\frac{\rho _{n}}{\rho _{s}}\delta \mathbf{v}%
_{n},\ \ \ \ \ \ \ \ \mathbf{v}_{ns}=\mathbf{v}_{ns0}+\frac{\rho }{\rho _{s}}%
\delta \mathbf{v}_{n}.  \label{net flow}
\end{equation}%
Substituting Eqs. (\ref{linearization}) and (\ref{net flow}) into the Vinen
equation (\ref{VE 1}) we arrive at
\begin{equation}
\frac{\partial \delta \mathcal{L}}{\partial t}=\alpha _{V}\;\frac{\rho }{%
\rho _{s}}\mathcal{L}_{0}^{3/2}\delta \mathbf{v}_{n}-\beta _{V}\frac{1}{2}%
\mathcal{L}_{0}\delta \mathcal{L}.  \label{viaL0}
\end{equation}

Equation (\ref{viaL0}) shows, that the evolution of fluctuating part of the
vortex line density $\delta \mathcal{L}$ bears the relaxation-type character
with a characteristic time $\tau _{V}=2/\left( \beta _{V}\mathcal{L}%
_{0}\right) $ and with a "force" proportional to $\delta \mathbf{v}_{n}$. It
allows to express the spectrum of VLD $\left\langle \delta \mathcal{L}%
(\omega )\delta \mathcal{L}(-\omega )\right\rangle $ via the spectrum of a
normal velocity. We accept further the usual in theory of turbulence
relationship $k=\omega /v_{n}$ between the wavenumber $k$\ and the frequency
$\omega $\ ($v_{n}$ is the mean flow).\ In the Fourier transforms Eq. (\ref%
{viaL0}) takes a form \

\begin{equation}
i\omega \delta \mathcal{L}(\omega )=\alpha _{V}\;\frac{\rho }{\rho _{s}}%
\mathcal{L}_{0}^{3/2}\delta \mathbf{v}_{n}(\omega )-\beta _{V}\frac{1}{2}%
\mathcal{L}_{0}\delta \mathcal{L}(\omega ),  \label{L in FT}
\end{equation}%
therefore the spectrum $\left\langle \delta \mathcal{L}(\omega )\delta
\mathcal{L}(-\omega )\right\rangle $ is
\begin{equation}
\left\langle \delta \mathcal{L}(\omega )\delta \mathcal{L}(-\omega
)\right\rangle =\frac{4(\alpha _{V}/\beta _{V})^{2}\;(\frac{\rho }{\rho _{s}}%
)^{2}\mathcal{L}_{0}\left\langle \delta \mathbf{v}_{n}(\omega )\delta
\mathbf{v}_{n}(-\omega )\right\rangle }{1+(\omega \tau _{V})^{2}}.
\label {spectrum VE}
\end{equation}%
Relation (\ref{spectrum VE}) shows that for small frequencies, $\omega
<1/\tau $, the spectrum of the VLD reproduces the spectrum of fluctuations
of the normal component, and if the Kolmogorov-type turbulence is developed
in the normal component, then quantity $\left\langle \delta \mathcal{L}%
(\omega )\delta \mathcal{L}(-\omega )\right\rangle $ scales as $\omega
^{-5/3}$. Interestingly, if we accept the conditions of the experiment by
Roche et al.\cite{Roche2007}, then (e.g. for the mean flow $v_{n}\approx 1$\
m/s at $T=1.6$\ K) we have
\begin{equation*}
\left\langle \delta \mathcal{L}(\omega )\delta \mathcal{L}(-\omega
)\right\rangle \approx 4\ast 10^{22}\omega ^{-5/3},
\end{equation*}%
which is close to the experimental data in order of magnitude. Note also the
dependence on the applied velocity (about $\varpropto v_{n}^{4}$) is also
consistent with experimental data.

\section{HVBK\ case}

In the case of quasi-classical turbulence, the set of vortex line is
believed to consist of the many bundles containing a large number of threads
inside of them. The macroscopic behavior of these bundles is quite similar
to the dynamics of eddies in ordinary fluids. The coarse-grained
hydrodynamic of the vortex bundles is studied by many authors (see e.g.,
\cite{Sonin1987},\cite{HENDERSON2000},\cite{Holm2001},\cite{Jou2011}),

but the basis of these stidies was the hydrodynamics of rotating
superfluids, or the Hall-Vinen-Bekarevich-Khalatnikov (HVBK) model (see
e.g., book \cite{Khalatnikov1965}). In the vortex bundles, the vorticity
field of $\omega _{s}$ (to be exact, its absolute value) and the vortex line
density $\mathcal{L}$ are related to each other with the use of the
Feynman's rule:

\begin{equation}
\omega _{s}=\kappa \mathcal{L}.  \label{bundle omega}
\end{equation}%
This formula reflets the fact that the VLD $\mathcal{L}$, in this case
coincides with the areal density of lines in a plane perpendicular to the
bundle. In terms HVBK dynamics of the vorticity obeys to the following
equation (see \cite{Khalatnikov1965})
\begin{equation}
\frac{\partial \mathbf{\omega }_{s}}{\partial t}=\nabla \times \lbrack
\mathbf{v}_{L}\times \mathbf{\omega }_{s}],  \label{dw/dt HVBK}
\end{equation}%
where $\mathbf{v}_{L}$ is the velocity of lines,

\begin{equation}
\mathbf{v}_{L}\mathbf{=v}_{s}\mathbf{+}\alpha \left[ \mathbf{\hat{\omega}}%
_{s}{\times }\left( \mathbf{v}_{n}-\mathbf{v}_{s}\right) \right] .
\label{V_L}
\end{equation}%
The meaning of Eq. (\ref{dw/dt HVBK}) is that it describes the motion of
vortex lines in the transverse (with respect to the unit vector $\mathbf{%
\hat{\omega}}_{s}$ along the vorticity) direction, when the coarse-grained
superfluid velocity $\mathbf{v}_{s}$ differs from the normal velocity $%
\mathbf{v}_{n}$. This equation is derived again without fluctuations. To
take into account the latter, we use the formula (\ref{linearization}) into
the equation (\ref{dw/dt HVBK}) and have%
\begin{equation*}
\frac{\partial \mathbf{\hat{\omega}}_{s}\mathbf{(}\mathcal{L}+\delta
\mathcal{L)}}{\partial t}=\nabla \times \lbrack (\mathbf{v}_{ns}+\delta
\mathbf{v}_{ns})(\mathcal{L}+\delta \mathcal{L})].
\end{equation*}%
After little algebra this equations is transformed to
\begin{equation}
\frac{\partial \mathbf{\hat{\omega}}_{s}\mathbf{(}\delta \mathcal{L)}}{%
\partial t}=\alpha \mathcal{L}\nabla \times \delta \mathbf{v}_{ns}
\label{dL HVBK}
\end{equation}%
The HVBK theory describes the redistribution of preexisting vortex lines in
the transverse direction, it does not include a mechanism of the appearance
of new lines. In papers \cite{Samuels1993},\cite{Barenghi1997},\cite%
{Barenghi1999} it was shown that the bundle structure of quantized vortices
develops inside the eddies of the normal component due to proliferation of
vortex filaments. Mechanism of this proliferation is quite involved, it
reminds the developing of the vortex tangle in an applied counterflow with
the growth of the number lines due to reconnections, and with the growth of
length due to relative velocity with the subsequent polarization. We take
into account this process by adding a relaxation-type term in Eq. (\ref{dL
HVBK}):

\begin{equation}
\frac{\partial \mathbf{\hat{\omega}}_{s}\mathbf{(}\delta \mathcal{L)}}{%
\partial t}=\alpha \mathcal{L}\nabla \times \delta \mathbf{v}_{ns}+\frac{1}{%
\tau _{s}}\delta \mathcal{L}.  \label{dL HVBK with Samuels}
\end{equation}%
Here $\tau _{s}$ is the time of relaxation of the bundle structure as
described by Samuels et al. \cite{Samuels1993},\cite{Barenghi1997},\cite%
{Barenghi1999}. In Fourier the component we have equation
\begin{equation*}
-i\omega \delta \mathcal{L}(\omega )=i\alpha \mathcal{L}\frac{\rho }{\rho
_{s}}(\mathbf{k}\times \delta \mathbf{v}_{ns})+\frac{1}{\tau _{s}}\delta
\mathcal{L},
\end{equation*}%
which lead to the following spectral density:

\begin{equation}
<\delta \mathcal{L}(\omega )\;\delta \mathcal{L}(-\omega )>=\frac{\alpha ^{2}%
\mathcal{L}^{2}(\frac{\rho }{\rho _{s}})^{2}\tau _{s}^{2}k^{2}\left\langle
\delta \mathbf{v}_{n}(\omega )\delta \mathbf{v}_{n}(-\omega )\right\rangle }{%
1+(\omega \tau _{s})^{2}}  \label{spectrum HVBK}
\end{equation}%
Again, as in the case of Vinen turbulence, the shape of the spectrum depends
on the value of $\omega \tau _{s}$. The question of the relaxation time is a
bit vague and unclear. Analyzing the numerical results, Samuels \cite%
{Samuels1993} offered an expression for $\tau _{s}$, which included the
circulation of the vortex tube (of normal component) and its size. Therefore
it is impossible to apply directly his result. The important fact is that
quantity $\tau _{s}$ is proportional $1/\sqrt{\alpha }$, thus, it decreases
with increasing temperature.

For large $\tau _{s}$, which implies a small coupling between the normal and
superfluid components and is realized for a small temperature, the spectrum $%
<\delta \mathcal{L}(\omega )\;\delta \mathcal{L}(-\omega )>\varpropto
\left\langle \delta \mathbf{v}_{n}(\omega )\delta \mathbf{v}_{n}(-\omega
)\right\rangle $, ($\omega ^{-5/3}$ for Kolmogorov turbulence in the normal
component). In this case the result is similar to the previous (Vinen
turbulence) case, considered above. Even the value of spectrum is close to
the one, which is induced by the counterflowing turbulence. However, for
small $\omega \tau _{s}<<1$, which corresponds to large temperature (strong
coupling due to the large mutual friction) the spectrum behaves as $<\delta
\mathcal{L}(\omega )\;\delta \mathcal{L}(-\omega )>\varpropto \omega
^{2}\left\langle \delta \mathbf{v}_{n}(\omega )\delta \mathbf{v}_{n}(-\omega
)\right\rangle $, \ ($\omega ^{1/3}$ for Kolmogorov turbulence), and the
intensity of this spectrum is much lower. This result is in good qualitative
agreement with the numerical result \cite{Salort2011},\cite{Baggaley2012a},%
\cite{Baggaley2011c}). As for the quantitative analysis, there is problem of
determination of the relaxation time $\tau _{s}$, whivh is beyond the scope
of the present study.

\section{Discussion and conclusion}

We studied \ analytically the spectrum of fluctuations of the vortex line
density both in the counterflowing (Vinen)\ turbulence and on the basis of
the HVBK theory. In both case these deviations of quantity $\delta \mathcal{L%
}(t)$ appear due to strongly fluctuating field of the normal velocity.
Deviations of $\delta \mathcal{L}(t)$ evolve in the relaxation like manner,
which is determined either from the Vinen equation or from the HVBK \
equation. It allows to find Fourier transform and evaluate the spectra $%
<\delta \mathcal{L}(\omega )\;\delta \mathcal{L}(-\omega )>$. The crucial
difference between these two cases is that the stirring force for $\mathcal{L%
}(t)$ in the Vinen case is proportional to perturbations of the normal
velocity $\delta \mathbf{v}_{n}(t)$, whereas in the HVBK this force is
related to $\nabla \times \delta \mathbf{v}_{n}(t)$. This difference results
in different spectra and their dependence on the temperature. From analysis
of the final relations for spectral densities $<\delta \mathcal{L}(\omega
)\;\delta \mathcal{L}(-\omega )>$ we can conclude that for the small
temperature, both model offer the $\omega ^{-5/3}$ specrtrum, whereas for
large temperature the HVBK theory results in the $\omega ^{1/3}$ dependence.
This conclusion may serve as a basis for  experimental determination of what
kind of turbulence is realized in different types of generation.

\section*{ACKNOWLEDGMENTS}

The work was supported by the grants N 10-08-00369 and N 10-02-00514 from
the Russian Foundation of Basic Research, and by the grant from the
President Federation on the State Support of Leading Scientific Schools
NSh-6686.2012.8.

\end{document}